\documentclass[letterpaper, 10 pt, conference]{ieeeconf}  

\IEEEoverridecommandlockouts                              

\usepackage{graphics} 
\usepackage{epsfig} 
\usepackage{mathptmx} 
\usepackage{times} 
\usepackage{amsmath} 
\usepackage{amssymb}  

\usepackage{bm}

\title{\LARGE \bf
Design of Resonance Ratio Control with Relative Position Information for Two-inertia System
}

\author{Kenta Araake$^{1}$, Sho Sakaino$^{2}$ and Toshiaki Tsuji$^{3}$
\thanks{$^{1}$Kenta Araake is a student with the Graduate School of Science and Engineering,
Saitama University, 255 Shimo-Okubo, Sakura-ku, Saitama City, Saitama 338-8570, Japan
        {\tt\small email: k.araake.362@ms.saitama-u.ac.jp}}%
\thanks{$^{2}$Sho Sakaino is with the Graduate School of Systems and Information Engineering, University of Tsukuba, 1-1-1 Tennodai, Tsukuba, Ibaraki 305-8577, Japan and the JST PRESTO
        {\tt\small email: sakaino@iit.tsukuba.ac.jp}}%
\thanks{$^{3}$Toshiaki Tsuji is with the Graduate School of Science and Engineering, 
Saitama University, 255 Shimo-Okubo, Sakura-ku, Saitama City, Saitama 338-8570, Japan
        {\tt\small email: tsuji@ees.saitama-u.ac.jp}}%
}

\begin{document}

\maketitle
\thispagestyle{empty}
\pagestyle{empty}

\begin{abstract}
Two-inertia systems are prone to resonance vibrations that degrade their control performances. These unwanted vibrations can be effectively suppressed by control methods based on a disturbance observer (DOB). Vibration suppression control methods using the information of both the motor and load sides have been widely researched in recent years. Methods that exploit the spring deflection or torsional force of two-inertia systems have delivered promising performances. However, few conventional methods have exploited the relative position information, and the discussion of position control is currently insufficient. Focusing on the relative position, this study proposes a new resonance ratio control (RRC) based on the relative acceleration and state feedback. The structure of the proposed RRC is derived theoretically and the proposed method is experimentally validated.

\end{abstract}

\section{Introduction}
\begin{figure}[t]
\centerline{\includegraphics[width=6.0cm]{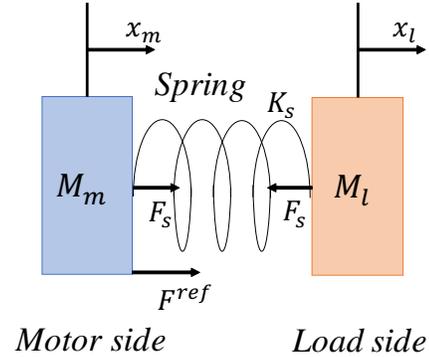}}
\caption{Model of a two-inertia system}
\label{two-inertia}
\end{figure}

High-performance position or velocity control has advanced alongside the development of actuators, central processing units, and sensors. Despite of the large progress of these technologies, the resonance problem remains a major problem because resonance vibrations (which depend on the mechanical system) degrade the control performance. Resonance is especially strongly in robots with flexible joints, in which the actuators and loads are connected by elastic components. Those systems are often modeled as two-inertia systems shown in Fig.~\ref{two-inertia}. Typical examples are timing-belt systems \cite{belt}, series elastic actuators \cite{sea}, \cite{sea2}, hydraulic actuators \cite{hyd}, and many of robot joints with gears \cite{gear}. Therefore, improving the control performance of two-inertia systems has been the subject of many researches \cite{a} \cite{b} \cite{c}.


Many of the existing vibration suppression controllers for two-inertia systems exploit the robustness of a disturbance observer (DOB) \cite{DOB} \cite{oboe}. The authors of \cite{oboe} improved the robustness of a controller by installing an accelerometer at the load side. Yuki {\it et al.} \cite{RRC} proposed a method for vibration suppression by resonance ratio control (RRC) based on a DOB. The RRC increases the resonance frequency by lowering the mass of the motor side of the two-inertia system. However, the RRC is not robust against disturbances because the load side information is estimated by a state observer. Another effective control against resonance is self-resonance cancellation (SRC), which cancels the resonance using the center of gravity of the motor and load sides of a two-inertia system \cite{SRC}. Self-resonance cancellation DOB (SRCDOB), which uses the information of the motor and the load sides, promises to solve the robustness problem. However, both SRC and SRCDOB require the accurate identification of many parameters. Any identification errors deteriorate the control performance. Other effective vibration suppression methods are full-state feedback controllers \cite{full}. Full-state feedback controllers use the information of both the motor and the load sides. The full-state feedback control proposed in \cite{full} employs a full-state feedback controller based on a new state equation using the relative velocity between the motor and load sides, without requiring the motor side position. Consequently, this design achieves stable position control even in systems with angular drift.

Several recent control methods exploit the spring deflection or torsional force of a two-inertia system. A DOB considering the spring deflection is more robust against modeling errors. It also delivers better force control of a two-inertia system than general DOBs using the information of the motor sides \cite{Oh}. The authors of \cite{Oh2} proposed a reduced-order DOB (RODOB) that exploits the spring deformation of two-inertia systems, and captures only the resonance frequency. The RODOB method requires fewer parameters than general DOBs, enabling easier system identification. In addition, high robustness and precise force control has been reported in an RRC with a torsional torque sensor, which monitors the torsional force in two-inertia systems \cite{Yokokura}. The control methods in \cite{Oh}, \cite{Oh2}, and \cite{Yokokura} use the relative position information, but position control has not been discussed. To address this deficiency, this paper proposes a new RRC with relative position information and a state feedback controller. Specifically, we integrate RODOB with RRC for robust position control. The proposed method is experimentally validated.

The remainder of this paper is structured as follows. Section~II, describes model of a two-inertia system. The new RRC with relative position information is proposed in section~III. The proposed method is experimentally validated and discussed in section~IV. Section~V concludes the paper.




\section{two-inertia system}
Many modern industrial robots and mechatronics systems transmit forces through a series of gears or belts between the motors and loads. Especially, if the transmitters between the motors and loads are flexible structures, they are modeled as two-inertia systems, which are liable to degradation of control performance. A two-inertia system model is shown in Fig.~\ref{two-inertia}.

The system consists of a motor mass $M_m$ (the input edge) and a load mass $M_l$ (the output edge) connected by a flexible structure modeled as a spring. $F^{ref}$ and $F_s$ represent reference motor force and spring force, respectively.

\begin{figure}[t]
\centerline{\includegraphics[width=8.0cm]{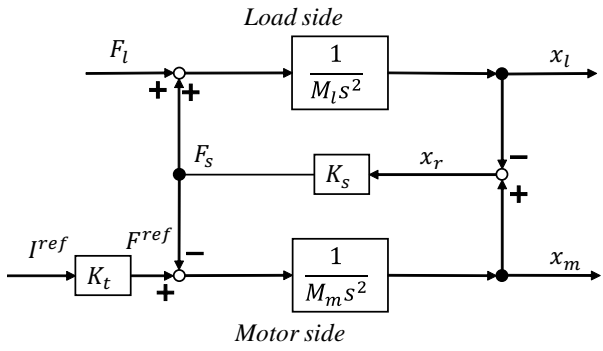}}
\caption{Block diagram of the two-inertia system}
\label{two-inertia block}
\end{figure}

Denoting the mass and position as $M_m$ and $x_m$ respectively on the motor side and $M_l$ and $x_l$ respectively on the load side, and representing the spring coefficient $K_s$, the equations are given by
\begin{eqnarray}
M_m\ddot{x}_m &=& -K_s(x_m - x_l) + F^{ref} - F_{m}^{dis}\label{1} \label{M} \\
M_l\ddot{x}_l &=& K_s(x_m - x_l) - F_{l}^{dis} \label{2}\label{L}
\end{eqnarray}
where $F_{m}^{dis}$ and $F_{l}^{dis}$ represent disturbance force of the motor and load side, respectively, and the dampers of the systems are ignored to simplify the discussion.
Fig.~\ref{two-inertia block} is a block diagram of the system, $I^{ref}$ and $F^{ref}$ are the current and force of the inputs on the motor side, respectively, $K_t$ is the force coefficient of the motor, and $F_l$ is the external force on the load side. The relative position $x_r$ is defined as follows:
\begin{eqnarray}
x_r&=&x_m - x_l\\
F_s &=& K_s x_r
\end{eqnarray}
where $F_s$ represents the spring force.

From Fig.~\ref{two-inertia block}, the transfer functions from the force reference $F^{ref}$ to the position $x_m$ at the motor side, and from $F^{ref}$ to the position $x_l$ at the load side are respectively calculated as follows:
\begin{eqnarray}
\frac{x_m}{F^{ref}} &=& \frac{1}{M_m s^2}\frac{s^2 + \omega_z^2}{s^2 + \omega_p^2} \label{XM} \\
\frac{x_l}{F^{ref}} &=& \frac{1}{M_m s^2}\frac{\omega_z^2}{s^2 + \omega_p^2} \label{XL}
\end{eqnarray}
where $\omega_p$ and $\omega_z$ respectively denote the resonance and antiresonance frequencies as follows:
\begin{eqnarray}
\omega_p &=& \sqrt{K_s\left( \frac{1}{M_m} + \frac{1}{M_l} \right)}\\
\omega_z &=& \sqrt{\frac{K_s}{M_l}}.
\end{eqnarray}

\section{design of the proposed controller}
\subsection{Resonance ratio control}
\begin{figure}[t]
\vspace{0.1in}
\centerline{\includegraphics[width=8.0cm]{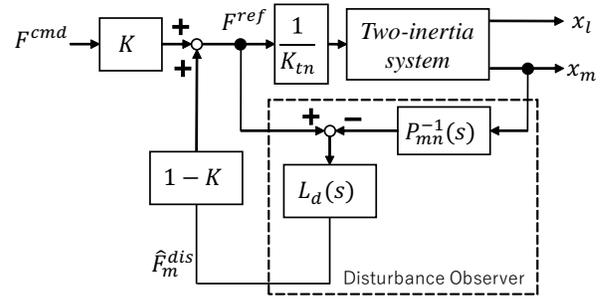}}
\caption{Block diagram of standard resonance ratio control}
\label{reso block}
\end{figure}

One known vibration suppression control is resonance ratio control (Fig.~\ref{reso block}), which consists of a feedback controller and a DOB based on resonance frequency of the two-inertia system \cite{RRC}. Here, $F^{cmd}$ is the force command and $K$ is the RRC gain, which must be properly designed. The nominal dynamics $P_{mn}(s)$ on the spring-less motor side are given by
\begin{equation}
P_{mn} (s) = \frac{1}{M_{mn} s^2}
\end{equation}
where the subscript $n$ means the nominal value.
$L_d(s)$ is the low pass filter (LPF) used in the DOB and expressed as $L_d(s) = g/(s+g)$, and $g$ is the cut-off frequency of the LPF.
DOBs generally emphasize the resonance in two-inertia systems by stiffening the motor side, including vibration at load side. Under RRC, the estimated disturbance $\hat{F}_m^{dis}$ calculated by the DOB is fed back by a factor of $1-K$. Meanwhile the reference force $F^{ref}$ is multiplied by $K$ before inputting to the motor. The motor mass is thus reduced by a factor of  $1/K$, severely suppressing the induced vibrations.

\subsection{Proposed resonance ratio control based on relative position}
This proposed control method uses the relative position $x_r$, represented as $x_r = x_m -x_l$. The transfer function $P_r(s)$ from the force reference $F^{ref}$ to the relative position $x_r$ is given by
\begin{equation}
P_r(s) = \frac{x_r}{F^{ref}} = \frac{1}{M_m(s^2+\omega_p^2)}.
\end{equation}
As mentioned above, two-inertia systems based on the relative position are regarded as second-order systems. 
Here, considering a case where disturbances occur on the motor and load side.  Eqs.~\eqref{M}, \eqref{L}, \eqref{XM}, and \eqref{XL} derives the motor position $x_m$ as follow:
\begin{equation}
x_m = \frac{1}{M_{m} s^2} \frac{1}{s^2+\omega_{p}^{2}}\{(s^2+\omega_{z}^{2}) (F^{ref} - F_{m}^{dis}) -\omega_{z}^2 F_{l}^{dis}\}. \label{XMdis}
\end{equation}
On the other hand, the relative position $x_r$ represent as Eq.~\eqref{XRdis} by Eqs.~\eqref{M}, \eqref{L}, \eqref{XM}, and \eqref{XL} when disturbances occur on the motor and load side.
\begin{equation}
x_r = \frac{1}{M_m(s^2 + \omega_{p}^{2})}(F^{ref} - F_{m}^{dis} + \frac{M_m}{M_l}F_{l}^{dis}) \label{XRdis}
\end{equation}
Eqs.~\eqref{XMdis} and \eqref{XRdis} show that the disturbances are affected by the fourth-order system, in the motor position; however, in the relative position, the disturbances are affected by the second-order system. Therefore, focusing on the relative position leads to suppressing the influences of disturbance and raising higher cutoff frequency of DOB.

The DOB in the proposed method counteracts the disturbance caused by the spring force. Fig.~\ref{proposed_block} is a block diagram of the proposed RRC. Here, $\hat{F_r}$ represents the estimated spring force, including the disturbances. As shown in the figure, only one parameter, $M_m$, must be identified, which greatly simplifies the method. The RRC essentially modifies the two-inertia system governed by \eqref{M} and \eqref{L} into a different system with the following state equations:


\begin{figure}[t]
\centerline{\includegraphics[width=8.0cm]{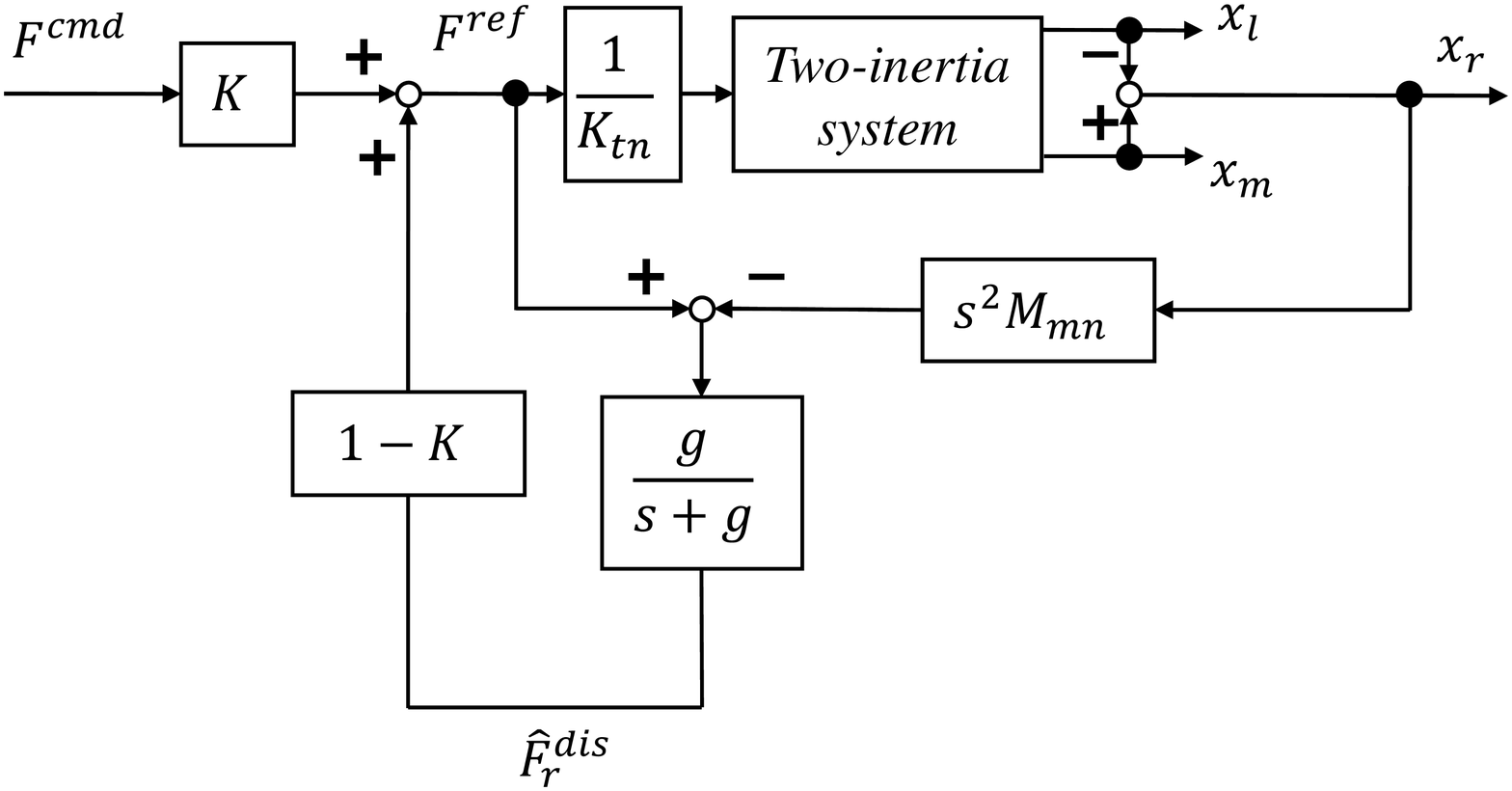}}
\caption{Block diagram of the proposed method}
\label{proposed_block}
\end{figure}

\begin{eqnarray}
\dot{\bm{x}}&=&\bm{Ax}+\bm{b}u\label{state} \\
\bm{y}&=&\bm{cx}\\
\bm{A} &=& \left[
\begin{array}{cccc}
0 & 1 & 0 & 0\\
-\frac{{K}_{s}^{\prime}}{{M_{m}^{\prime}}} & 0 & \frac{K_{s}^{\prime}}{M_{m}^{\prime}} & 0\\
0 & 0 & 0 & 1\\
\frac{K_{s}^{\prime}}{M_{l}^{\prime}} & 0 & -\frac{K_{s}^{\prime}}{M_{l}^{\prime}} & 0
\end{array}
\right]\\
\bm{b} &=& \left[
\begin{array}{cccc}
0 & \frac{1}{M_{m}^{\prime}} & 0 & 0
\end{array}
\right]^{\mathrm{T}}\\
\bm{x} &=& \left[
\begin{array}{cccc}
x_{m} & \dot{x}_{m} & x_{l} & \dot{x}_{l}
\end{array}
\right]^{\mathrm{T}}\\
\bm{c} &=& \left[
\begin{array}{cccc}
0 & 0 & 1 & 0
\end{array}
\right]
\end{eqnarray}
where $M_{m}^{\prime}$, $M_{l}^{\prime}$, and $K_{s}^{\prime}$ denote the motor mass, the load mass, and the spring coefficient modified by the proposed RRC, respectively. These parameters are respectively calculated as follows: 
\begin{eqnarray}
M_{m}^{\prime} &=& \frac{M_m}{K}\label{M_m2} \\
M_{l}^{\prime} &=& \frac{K(M_m+M_l)-M_m}{K} \label{M_l2} \\
K_{s}^{\prime} &=& \frac{K(M_m+M_l)-M_m}{KM_l}K_{s} \label{K_s2}.
\end{eqnarray}
As described above, the motor mass $M_m$ in the proposed RRC is only $1 / K$ times the original motor mass. However, as shown by Eqs.~\eqref{M_m2} and \eqref{M_l2}, the total mass $M_m + M_l = M_{m}^{\prime} + M_{l}^{\prime}$ is unchanged.
Moreover, Eq.~\eqref{ks/ml} (derived from Eqs.~\eqref{M_l2} and \eqref{K_s2}) confirms that pole on the load side is also unmodified. Therefore, the resonance frequency of the modified system $\omega_{p}^{\prime}$ is expressed as follows:
\begin{eqnarray}
\frac{K_{s}^{\prime}}{M_{l}^{\prime}} &=& \frac{K_s}{M_l}\label{ks/ml} \\
\omega_{p}^{\prime} &=& \sqrt{K K_s\left( \frac{1}{M_m} + \frac{1}{M_l} \right)} = \sqrt{K}\omega_p.
\end{eqnarray}

\section{experiments and discussion}
\subsection{Experimental setup}
\begin{figure}[t]
\centerline{\includegraphics[width=8.0cm]{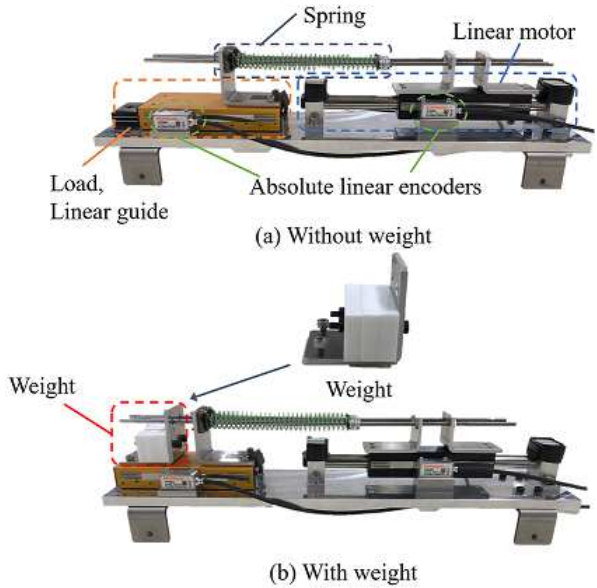}}
\caption{Experimental setup}
\label{setup}
\end{figure}

This subsection describes the experimental setup of the two-inertia system. The system comprises a linear actuator, a load, and a spring connecting the motor to the load (see Fig.~\ref{setup}).  The system can be equipped with a weight, as shown in Fig.~\ref{setup}~(b). The linear actuator, S160Q(GHC), produces a force up to 80 N. The positions of the motor and load were measured by absolute linear encoders with a resolution of 50~nm, and a control period was $0.1$~msec.

The parameters derived from the system responses to pseudo random binary signal inputs. Due to the limited movable range, the binary signals 0 and 1 were set to 10~N and 20~N respectively, or to -10~N and -20~N, respectively, as specified in \cite{iden}. By setting 10~N instead of 0~N as the low binary signal, we can ignore the static friction fo the system.
The identified system parameters obtained in the absence and presence of the weight are shown in Tables~\ref{parameter} and \ref{weight}, respectively. Noting that the identified values of the motor mass and the spring coefficient also changed when adding a weight.
\begin{table}[t]
\caption{System parameters without a weight}
\begin{center}
\begin{tabular}{|l||l|l|}
\hline
$K_t$&Force coefficient& 33.0 N/A\\
$M_m$&Mass (motor side)& 1.20 kg\\
$M_l$&Mass (load side)& 1.09 kg\\
$K_s$&Spring coefficient& 4662 N/m\\ \hline
$f_p$&Resonance frequency& 14.4 Hz\\
$f_z$&Antiresonance frequency& 10.4 Hz\\
\hline
\end{tabular}
\label{parameter}
\end{center}
\end{table}

\begin{table}[!t]
\vspace{0.1in}
\caption{System parameters with a weight}
\begin{center}
\begin{tabular}{|l||l|l|}
\hline
$M_m$&Mass (motor side)& 1.26 kg\\
$M_l$&Mass (load side)& 1.59 kg\\

$K_s$&Spring coefficient& 4917 N/m\\ \hline
$f_p$&Resonance frequency& 13.3 Hz\\
$f_z$&Antiresonance frequency& 8.85 Hz\\
\hline
\end{tabular}
\label{weight}
\end{center}
\end{table}

\subsection{Control gain}
The proposed method includes a newly designed RRC and an outer PD loop for state feedback control. This subsection describes the gain of the implemented control. Figure~\ref{whole_block} is the whole block diagram of the proposed method. Here, the subscripts $res$ and $cmd$ denote the response and command values, respectively. The gains $K_{pm}$ ($K_{dm}$) and $K_{pl}$ ($K_{dl}$) denote the proportional (derivative) gains of the motor and load, respectively. The four gains are represented as follows:

\begin{figure}[t]
\vspace{0.1in}
\centerline{\includegraphics[width=8.0cm]{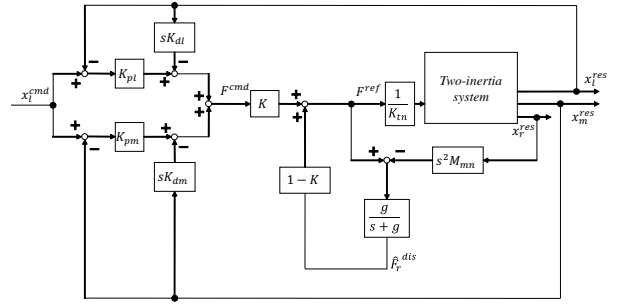}}
\caption{Whole block diagram of the proposed method}
\label{whole_block}
\end{figure}

\begin{eqnarray}
f &=& \left[
\begin{array}{cccc}
K_{pm} & K_{dm} & K_{pl} & K_{dl}
\end{array}
\right] \label{F} \\
u&=&-fx \label{FF}
\end{eqnarray}
where $u$ is the input force in Eq.~\eqref{state}. From Eqs.~\eqref{state}, \eqref{F}, and \eqref{FF}, the transfer function from the reference force to the load position $T(s)$ is derived as follows:
\begin{eqnarray}
T(s)=\frac{x_l^{res}}{F^{ref}}&=&\frac{\frac{K^2 K_s}{M_m M_l}}{s^4 + a_3s^3 + a_2s^2 + a_1s + a_0}\\
a_3&=&\frac{K K_{dm}}{M_m}\\
a_2&=&\frac{K(K_s M_m + M_l (K_s + K_{pm}))}{M_m M_l}\\
a_1&=&\frac{K K_s(K_{dm} + K_{dl})}{M_m M_l}\\
a_0&=&\frac{K K_s(K_{pm} + K_{pl})}{M_m M_l}.
\end{eqnarray}

In all experiments of this paper, we selected a quadruple pole on $\alpha$. The feedback gains were selected as follows:
\begin{eqnarray}
K_{pm}&=&\frac{6 \alpha^2 M_{m} M_{l} - K K_{s}(M_{m} + M_{l})}{K M_{l}}\\
K_{pl}&=&\frac{M_{m}(\alpha^4 M_{l}^2 - 6 \alpha^2 K_{s} M_{l}) + K K_{s}^2(M_{m} + M_{l})}{K M_{l} K_{s}} \\
K_{dm}&=&\frac{4 \alpha M_{m}}{K}\\
K_{dl}&=&\frac{M_{m}(4 \alpha^3 M_{l} - 4 \alpha K_{s})}{K K_{s}}.
\end{eqnarray}


\subsection{Experiments}
This subsection validates the proposed RRC in series of experiments. For exact comparisons with the conventional RRC, for controller of the conventional RRC was designed with full-state feedback (as in the proposed RRC), and the resonance frequency to be modified was the same in both RRCs. The pole on $\alpha$ of the state feedback was also identical in both RRCs. The control parameters of the conventional and proposed methods are shown in Tables~\ref{conv} and \ref{prop}, respectively. Because the disturbances in the conventional method have the forth-order characteristics as shown \eqref{XMdis}, we could not set sufficiently high DOB gain. However, because those in the proposed method have the second-order characteristics as shown in \eqref{XRdis}, we could set a DOB gain as far greater than the resonance frequency.

\begin{table}[t]
\vspace{0.1in}
\caption{Control parameters of the conventional method}
\begin{center}
\begin{tabular}{|l||l|l|}
\hline
$g_l$&Gain of pseudo differential& 3000 rad/s\\
$g_d$&Gain of DOB& 100 rad/s\\
$\alpha$&Quadruple pole & 90 rad/s \\
$K$&Resonance ratio gain & 4.40 \\ \hline
$f_{p}^{\prime}$&Resonance frequency modified by RRC& 23.3 Hz\\
$M_{m}^{\prime}$&Motor side mass modified by RRC& 0.273 kg\\
\hline
\end{tabular}
\label{conv}
\end{center}
\end{table}

\begin{table}[t]
\caption{Control parameters of the proposed method}
\begin{center}
\begin{tabular}{|l||l|l|}
\hline
$g_l$&Gain of pseudo differential& 3000 rad/s\\
$g_d$&Gain of DOB& 500 rad/s\\
$\alpha$&Quadruple pole & 90 rad/s \\
$K$&Resonance ratio gain & 2.62 \\ \hline
$f_{p}^{\prime}$&Resonance frequency modified by RRC& 23.3 Hz\\
$M_{m}^{\prime}$&Motor side mass modified by RRC& 0.458 kg\\
$M_{l}^{\prime}$&Load side mass modified by RRC& 1.83 kg\\
$K_{s}^{\prime}$&Spring coefficient modified by RRC& 7836 N/m\\
\hline
\end{tabular}
\label{prop}
\end{center}
\end{table}

\subsubsection{Reference responses}
\begin{figure}[t]
\centerline{\includegraphics[width=6.0cm]{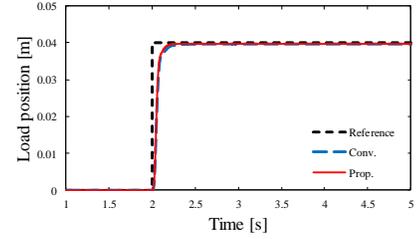}}
\caption{Step responses in the conventional (blue) and proposed (red) RRCs. The black dotted response is the reference step response}
\label{step}
\end{figure}

The step reference responses under the conditions of Fig.~\ref{setup}~(a) are shown in Fig.~\ref{step}. No significant differences between the conventional and proposed RRCs are evident. When the influences of the disturbances (including the modeling error) were small, the control performances of the two RRCs were almost identical.

\subsubsection{Parameter Variation}
\begin{figure}[t]
\centerline{\includegraphics[width=6.0cm]{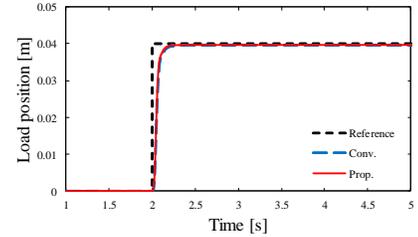}}
\caption{Step responses under the condition of $M_{mn} = 0.5M_m$}
\label{0.5Mm}
\end{figure}

\begin{figure}[t]
\centerline{\includegraphics[width=6.0cm]{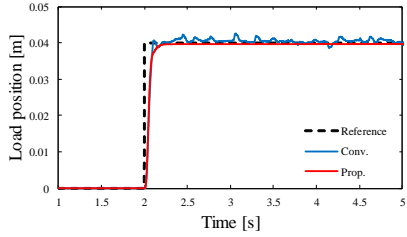}}
\caption{Step responses under the condition of $M_{mn} = 1.5M_m$}
\label{1.5Mm}
\end{figure}

To confirm the robustness against parameter variations, the step responses were monitored for different values of the control parameters. Figures~\ref{0.5Mm} and \ref{1.5Mm} show the results of multiplying the original motor mass by 0.5 and 1.5 times, respectively. The responses of both RRCs were robust to having the 0.5~times modeled motor mass (Fig.~\ref{0.5Mm}). However, when the modeled motor mass was increased 1.5 times, the response of the conventional RRC developed oscillations while the proposed RRC remained stable (Fig.~\ref{1.5Mm}). The results confirm that the proposed RRC was highly robust against modeling errors, and well suppressed the vibrations at the load side position owing to the second-order characteristics of the disturbances and the high DOB gain.

\subsubsection{Load weight variation}
\begin{figure}[t]
\centerline{\includegraphics[width=6.0cm]{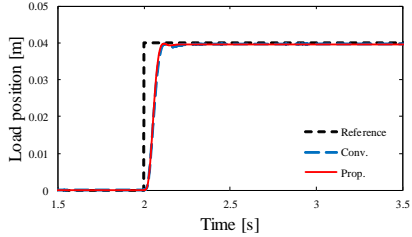}}
\caption{Step responses under a different load weight}
\label{added_Ml}
\end{figure}

To clarify the effects of variation of varying the load weight, we monitored the step responses under the conditions of Fig.~\ref{setup}~(b) and Table~\ref{weight} for different masses on the load side. The results are shown in Fig.~\ref{added_Ml}. Here, the control parameters (state feedback gain and motor mass of the DOB) were unchanged from those of the previous experiments, so the result confirms the robustness to load mass variations alone. Comparing Figs.~\ref{added_Ml} and \ref{step}, both RRCs developed slight vibrations under the higher load mass, but both were strongly robustness against load weight variations.

\subsubsection{Chirp responses}
\begin{figure}[t]
\centerline{\includegraphics[width=6.0cm]{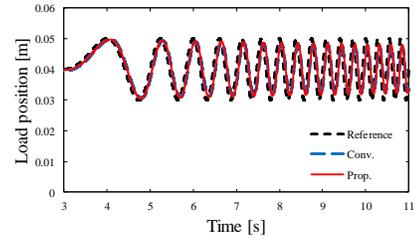}}
\caption{Chirp responses}
\label{chirp}
\end{figure}

\begin{figure}[t]
\centerline{\includegraphics[width=6.0cm]{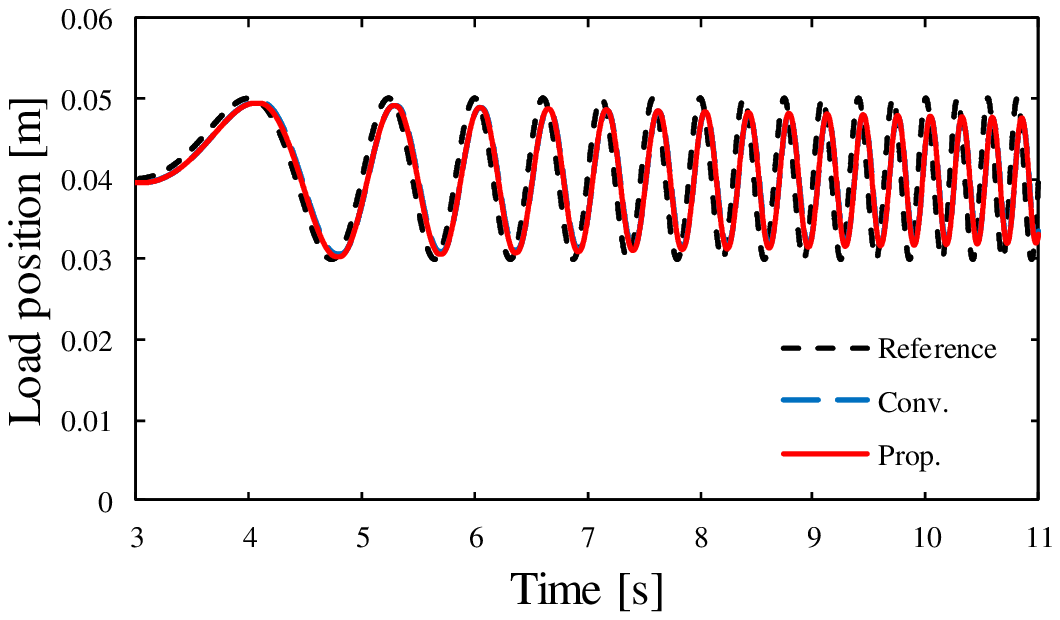}}
\caption{Chirp responses under the condition of $M_{mn} = 0.5M_m$}
\label{chirp_0.5Mm}
\end{figure}

\begin{figure}[!t]
\centerline{\includegraphics[width=6.0cm]{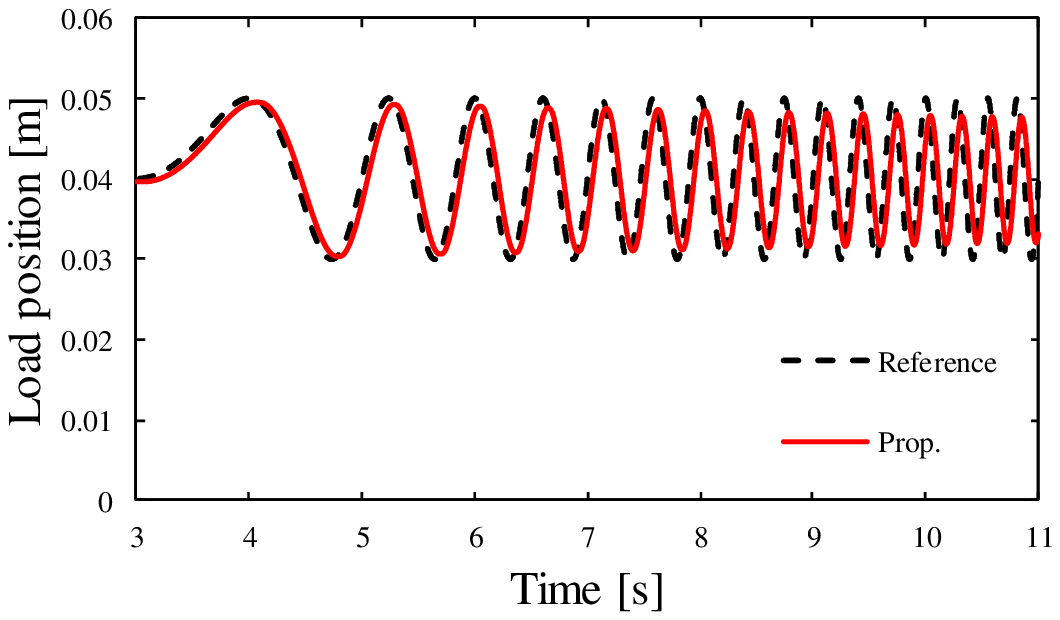}}
\caption{Chirp response under the condition of $M_{mn} = 1.5M_m$}
\label{chirp_1.5Mm}
\end{figure}

\begin{figure}[!t]
\centerline{\includegraphics[width=6.0cm]{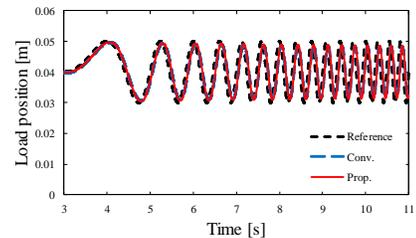}}
\caption{Chirp responses under a different load weight}
\label{chirp_added_Ml}
\end{figure}

To confirm the responses against frequency variations, the chirp signal responses were monitored. Figure~\ref{chirp} shows the result using nominal motor mass, and no significant differences between the conventional and proposed RRCs are evident. The results of multiplying the original motor mass by 0.5 and 1.5 times, respectively, are shown in Figs.~\ref{chirp_0.5Mm} and \ref{chirp_1.5Mm}. The both RRCs showed the robustness in Fig.~\ref{chirp_0.5Mm}; however, when the modeled motor mass was multiplied 1.5~times, the result of the conventional RRC was not obtained because the motor went out of control and could not move. On the other hand, the response of the proposed RRC was robust to having 1.5~times modeled motor mass as shown in Fig.~\ref{chirp_1.5Mm}. Figure~\ref{chirp_added_Ml} shows that both RRCs have the robustness against load mass vibrations as the step responses in Fig.~\ref{added_Ml}.

\section{conclusion}
This study proposed a new RRC that uses the relative position information and state feedback. This method alters the dynamics of the standard two-inertia system. The structure of the proposed RRC was theoretically derived, and was confirmed to tolerate higher cutoff frequencies of DOBs than standard two-inertia systems. The state feedback gains were also theoretically derived. Although the conventional RRC using state feedback provides excellent responses under many conditions, the proposed RRC (unlike the conventional design) ensures robustness against disturbances in modeling error, owing to the high cutoff frequency of its DOB. The experimental results confirmed the validity of the proposed RRC.

\section*{ACKNOWLEDGMENT}
This work was supported by the JST PRESTO Grand Number JPMJPR1755, Japan.


\begin{thebibliography}{00}
\bibitem{belt}A. Hasegawa, H. Fujimoto, and T. Takahashi, ``Robot joint angle control based on Self Resonance Cancellation using double encoders,'' in 2017 IEEE International Conference on Advanced Intelligent Mechatronics (AIM), Munich, Germany, 2017, pp. 460-465.
\bibitem{sea}G. A. Pratt and M. M. Williamson, ``Series elastic actuators,'' in Proceedings 1995 IEEE/RSJ International Conference on Intelligent Robots and Systems. Human Robot Interaction and Cooperative Robots, 1995, vol. 1, pp. 399-406 vol.1.
\bibitem{sea2}E. Sariyildiz, G. Chen, and H. Yu, ``An Acceleration-Based Robust Motion Controller Design for a Novel Series Elastic Actuator,'' IEEE Transactions on Industrial Electronics, vol. 63, no. 3, pp. 1900-1910, Mar. 2016.
\bibitem{hyd}T. Sakuma, K. Tsuda, K. Umeda, S. Sakaino, and T. Tsuji, ``Modeling and resonance suppression control for electro-hydrostatic actuator as a two-mass resonant system,'' Advanced Robotics, vol. 32, no. 1, pp. 1-11, Jan. 2018.
\bibitem{gear}Y. Hirano, T. Yoshioka, K. Ohishi, T. Miyazaki, Y. Yokokura, and M. Sato, ``Vibration Suppression Control Method for Trochoidal Reduction Gears under Load Conditions,'' IEEJ Journal of Industry Applications, vol. 5, no. 3, pp. 267-275, 2016.
\bibitem{a} Y. Hori, ``2-Inertia System Control using Resonance Ratio Control and Manabe Polynomials,'' IEEJ Trans. IA, vol. 114, no. 10, pp. 1038-1045, Sep. 1994.
\bibitem{b} S. Yamada, H. Fujimoto, and Y. Terada, ``Joint torque control for backlash compensation in two-inertia system,'' in 2016 IEEE 25th International Symposium on Industrial Electronics (ISIE), Santa Clara, CA, USA, 2016, pp. 1138-1143.
\bibitem{c} K. Tsuda, T. Sakuma, K. Umeda, S. Sakaino, and T. Tsuji, ``Resonance-suppression Control for Electro-hydrostatic Actuator as Two-inertia System,'' IEEJ Journal of Industry Applications, vol. 6, no. 5, pp. 320-327, 2017.
\bibitem{DOB} K. Ohnishi, ``Robust Motion Control by Disturbance Observer,'' JRSJ, vol. 11, no. 4, pp. 486-493, May 1993.
\bibitem{oboe}R. Oboe and D. Pilastro, ``Use of load-side MEMS accelerometers in servo positioning of two-mass-spring systems,'' in IECON 2015 - 41st Annual Conference of the IEEE Industrial Electronics Society, Yokohama, 2015, pp. 004603-004608.
\bibitem{RRC} K. Yuki, T. Murakami, and K. Ohnishi, ``Vibration Control of a 2 Mass Resonant System by the Resonance Ratio Control,'' IEEJ Trans. IA, vol. 113, no. 10, pp. 1162-1169, Oct. 1993.
\bibitem{SRC} M. Aoki, H. Fujimoto, Y. Hori, and T. Takahashi, ``Robust resonance suppression control based on self resonance cancellation disturbance observer and application to humanoid robot,'' in 2013 IEEE International Conference on Mechatronics (ICM), Vicenza, 2013, pp. 623-628.
\bibitem{full} S. Sakaino and T. Tsuji, ``Resonance Suppression of Electro-hydrostatic Actuator by Full State Feedback Controller Using Load-side Information and Relative Velocity,'' IFAC-PapersOnLine, vol. 50, no. 1, pp. 12065-12070, Jul. 2017.
\bibitem{Oh} S. Oh, C. Lee, and K. Kong, ``Force control and force observer design of series elastic actuator based on its dynamic characteristics,'' in IECON 2015 - 41st Annual Conference of the IEEE Industrial Electronics Society, Yokohama, 2015, pp. 004639-004644.
\bibitem{Oh2} H. Lee and S. Oh, ``Design of reduced order disturbance observer of series elastic actuator for robust force control,'' in 2018 IEEE 15th International Workshop on Advanced Motion Control (AMC), Tokyo, 2018, pp. 663-668.
\bibitem{Yokokura} Y. Yokokura and K. Ohishi, ``Single inertialization of a 2-inertia system based on fine torsional torque and sensor-based resonance ratio controllers,'' in 2017 IEEE International Conference on Mechatronics (ICM), Churchill, Australia, 2017, pp. 196-201.
\bibitem{iden}T. Yamazaki, S. Sakaino, and T. Tsuji, ``Estimation and Kinetic Modeling of Human Arm using Wearable Robot Arm,'' Electrical Engineering in Japan, vol. 199, no. 3, pp. 57-67, May 2017.

\end{thebibliography}
\end{document}